\RequirePackage{fix-cm}
\documentclass[twocolumn,epjc3]{svjour3}
\smartqed  % flush right qed marks, e.g. at end of proof
\RequirePackage{graphicx}
\begin{document}

\title{Semileptonic Transition of $\Lambda_{b}$ Baryon}


\author{Kaushal Thakkar}

\thankstext{e1}{e-mail: kaushal2physics@gmail.com}


\institute{Department of Physics, Government College, Daman-396210, U. T. of Dadra $\&$ Nagar Haveli and Daman $\&$ Diu, INDIA\\
}

\date{Received: date / Accepted: date}

\maketitle

\begin{abstract}
The semileptonic transition of $\Lambda_b$ baryon is studied using the Hypercentral constituent quark model. The six-dimensional hyperradial $Schr\ddot{o}dinger$ equation is solved in the variational approach to get masses and wavefunctions of heavy baryons. The matrix elements of weak decay are written in terms of the overlap integrals of the baryon wave function. The Isgur-Wise function is determined to calculate exclusive semileptonic decay  $\Lambda_b$ $\rightarrow$ $\Lambda_c$ $\ell$ $\bar{\nu}$. The calculated decay rate and the branching ratio of $\Lambda_b$ baryon are consistent with other theoretical predictions and with the available experimental observations.

\keywords{Bottom Baryons, Semileptonic Decays, Nonrelativistic quark model}
\end{abstract}

\section{Introduction}
\label{intro}

Inclusive and exclusive semileptonic decays of heavy flavour hadrons play an important role in the calculation of fundamental parameters of the electroweak standard model and towards a deeper understanding of QCD. The semileptonic decay of heavy hadrons is also a unique tool for determining the elements of the Cabibbo-Kobayashi-Maskawa (CKM) matrix, to study the internal structure of hadrons.

The semileptonic decay of heavy mesons have been studied extensively as mentioned in Refs. \cite{Bailey2014,Gambino2014,Benson2003,Gambino2012,Kapustin1996,Gambino2010} and references therein, but fewer attempts have been made to study the semileptonic decay of heavy baryons compare to that of heavy mesons. The chosen semileptonic $\Lambda_b$ $\rightarrow$ $\Lambda_c$ $\ell$ $\bar{\nu}$ transition is one of the prominent decay channels out of the manifold available channels of the $\Lambda_b$ baryon reported by PDG \cite{PDG2018}. This particular semileptonic transition has been investigated using different theoretical approaches such as Covariant Confined Quark Model \cite{Gutsche2015}, QCD Sum Rules \cite{{Azizi2018},Carvalho1999,{Huang2005}}, quark Model \cite{Pervin2005}, Bethe-Salpeter Equation \cite{Guo1996}, Lattice QCD \cite{Detmold2015,{Bowler1998}}, Zero recoil sum rules \cite{Mannel2015}, Relativistic Quark Model \cite{Faustov2016,{Ebert2006}},  Light Front Approach \cite{Hong-Wei2008} etc. Also the experimental group like DELPHI collaboration \cite{DELPHI Collaboration2004} and LHCb collaboration \cite{LHCb2017} reported their measurement on the slope parameter $\rho^2$ in the Isgur-Wise function and the branching ratio of the semileptonic process of $\Lambda_b$ baryon.

All these experimental measurements and theoretical calculations make the study of semileptonic decay of $\Lambda_b$ interesting. A precise calculation of form factors involved in the process of weak decay has been unrevealed for many years due to the perturbative nature of QCD. The Heavy-Quark Effective Theory (HQET) provides the framework to include non-perturbative corrections for studing hadrons containing heavy quarks. In the limit of infinite heavy-quark mass, all the form factors describing the semileptonic decay of heavy baryons are proportional to the universal function only, which is known as the Isgur-Wise (IW) function.

In this paper, we extend the study of our earlier work \cite{{Thakkar2017},{Majethiya2016},{ShahEPJA2016},{ShahCPC2016},{ShahEPJC2016}} on the mass spectra of heavy baryons to the study of exclusive semileptonic decay of $\Lambda_b$  baryon. This paper is organized as follows: The hypercentral Constituent Quark Model (HCQM) is applied to get masses and wave function of heavy baryons presented in section \ref{sec:2}. We have furnished a detailed calculation of Isgur-Wise function and the decay rate of semileptonic transition of $\Lambda_b$ baryon in section \ref{sec:3}. In section \ref{sec:4}, we have presented results and also drawn an important conclusion. Finally, the present study on the semileptonic transition of $\Lambda_b$ baryon is summarized in Section \ref{sec:5}.

\section{Hypercentral Constituent Quark Model (HCQM) for Baryons}\label{sec:2}
The exact solution of the QCD equations is very complex, so one has to rely upon conventional quark models. The assumptions in various conventional quark models are different, but they have a simple general structure in common including some basic features like confinement and asymptotic freedom and for the rest built up using suitable assumptions. In this article, we have adopted HCQM to study masses of heavy baryons ($\Lambda_c$, $\Lambda_b$) and semileptonic transition of $\Lambda_b$ baryon. For detailed information on Hypercentral Constituent Quark Model (HCQM), see references \cite{Ferraris1995,Giannini1983,Santopinto1995}.

The relevant degrees of freedom for the relative motion of the three constituent quarks are provided by the relative Jacobi coordinates $\vec{\rho}$  and $\vec{\lambda}$ which are given by \cite{Thakkar2017,{ShahEPJA2016}} as
\begin{equation}\label{eq:1}
\centering
\vec{\rho} = \frac{1}{\sqrt{2}}(\vec{r_{1}} - \vec{r_{2}})
\end{equation}
\begin{equation}\label{eq:2}
\vec{\lambda} =\frac{m_1\vec{r_1}+m_2\vec{r_2}-(m_1+m_2)\vec{r_3}}{\sqrt{m_1^2+m_2^2+(m_1+m_2)^2}}
\end{equation}
%\end{subequations}
%Here $m_i$ and $\vec{r_i}$ (i = 1, 2, 3) denote the mass and coordinate of the i-th constituent quark.
The respective reduced masses are given by
%\begin{subequations}
\begin{equation}\label{eq:3}
m_{\rho}=\frac{2 m_{1} m_{2}}{m_{1}+ m_{2}}
\end{equation}
\begin{equation}\label{eq:4}
 m_{\lambda}=\frac{2 m_{3} (m_{1}^2 + m_{2}^2+m_1m_2)}{(m_1+m_2)(m_{1}+ m_{2}+ m_{3})}
\end{equation}
%\end{subequations}
Here, $m_1$, $m_2$ and $m_3 $ are the constituent quark masses. The angles of the Hyperspherical coordinates are given by $\Omega_{\rho}= (\theta_{\rho}, \phi_{\rho})$ and $\Omega_{\lambda}= (\theta_{\lambda}, \phi_{\lambda})$. We define hyper radius $x$ and hyper angle $\xi$ by,
\begin{equation}\label{eq:5}
x= \sqrt{\rho^{2} + \lambda^{2}}\,\,\,and\,\,\, \xi= arctan \left(\frac{\rho}{\lambda} \right)
\end{equation}

In the center of mass frame ($R_{c.m.} = 0$), the kinetic energy operator can be written as
\begin{eqnarray}\nonumber\label{eq:6}
\frac{P_{x}^2}{2m}&&=-\frac{\hbar^2}{2m}(\bigtriangleup_{\rho} + \bigtriangleup_{\lambda})\\
 &&=-\frac{\hbar^2}{2m}\left(\frac{\partial^2}{\partial x^2}+\frac{5}{x}\frac{\partial}{\partial x}+\frac{L^2(\Omega)}{x^2}\right)
\end{eqnarray}
where $m$=$\frac{2 m_{\rho}m_{\lambda}}{m_{\rho}+m_{\lambda}}$ is the reduced mass and $L^2(\Omega)$
=$L^2$ $(\Omega_{\rho}$,$\Omega_{\lambda}$,$\xi)$ is the quadratic Casimir operator of the six-dimensional rotational group O(6) and its eigenfunctions are the hyperspherical harmonics, $Y_{[\gamma]l_{\rho}l_{\lambda}}$($\Omega_{\rho}$,$\Omega_{\lambda}$,$\xi$) satisfying the eigenvalue relation,
 $L^2Y_{[\gamma]l_{\rho}l_{\lambda}}$
($\Omega_{\rho}$,$\Omega_{\lambda},\xi)$=-$\gamma (\gamma +4) Y_{[\gamma]l_{\rho}l_{\lambda}}(\Omega_{\rho},\Omega_{\lambda},\xi)$. Here, $l_{\rho}$ and $l_{\lambda}$ are the angular momenta associated with the $\vec{\rho}$ and $\vec{\lambda}$
variables respectively and $\gamma$ is the hyper angular momentum quantum number.
\par The confining three-body potential is chosen within a string-like picture, where the quarks are connected by gluonic strings. And this potential increases linearly with a collective radius $x$ as mentioned in \cite{ginnani2015}. In the hypercentral approximation, the potential is expressed in terms of the hyper radius ($x$) as
\begin{equation}\label{eq:7}
\sum_{i<j}V(r_{ij})=V(x)+. . . .
\end{equation}
In this case, the potential $V(x)$ not only contains two-body interactions but it contains three-body
effects also. The three-body effects are desirable in the study of hadrons since the non-
Abelian nature of QCD leads to gluon-gluon couplings which produce three-body forces.

The  model Hamiltonian for baryons in the HCQM is then expressed as
\begin{equation}\label{eq:8}
H= \frac{P_{x}^{2}}{2m} +V(x)
\end{equation}

The six-dimensional hyperradial $Schr\ddot{o}dinger$ equation which corresponds to the above Hamiltonian can be written as
\begin{eqnarray}\nonumber\label{eq:9}
\left[\frac{d^2}{dx^2}+\frac{5}{x}\frac{d}{dx}-\frac{\gamma(\gamma+4)}{x^2}\right]&&\psi_{\nu\gamma}(x)= \\
&&-2m\left[E-V(x)\right]\psi_{\nu\gamma}(x)
\end{eqnarray}
%\left[\frac{-1}{2m}\frac{d^{2}}{d x^{2}} + \frac{\frac{15}{4}+ \gamma(\gamma+4)}{2mx^{2}}+ V(x)\right]\phi_{ \gamma}(x)= E\phi_{\gamma}(x)
%\begin{equation}\label{eq:9}
%\left[\frac{d^2}{dx^2}+\frac{5}{x}\frac{d}{dx}-\frac{\gamma(\gamma+4)}{x^2}\right]\psi_{\omega\gamma}(x)=-2m\left[E-V(x)\right]\psi_{\omega\gamma}(x)
%\end{equation}

where $\psi_{\nu\gamma}(x)$ is the  hyper-radial wave function. For the present study, we consider the hypercentral potential $V(x)$ as the hyper Coulomb plus linear potential which is given as

\begin{equation}\label{eq:10}
V(x)= \frac{\tau}{x}+ \beta {x}+V_0
\end{equation}

Here, the hyper-Coulomb strength is $\tau=-\frac{2}{3} \alpha_s$, where $\frac{2}{3}$ is the color factor for the baryon. The term $\beta$ corresponds to the string tension of the confinement. We fix the model parameters $\beta$ and $V_0$ to get the experimental ground state mass of $\Lambda_b$ baryon. The parameter $\alpha_s$ corresponds to the strong running coupling constant, which is written as
\begin{equation}\label{eq:11}
\alpha_{s}= \frac{\alpha_{s}(\mu_{0})}{1+\left(\frac{33-2n_{f}}{12 \pi}\right) \alpha_{s}(\mu_{0}) ln \left(\frac{m_{1}+ m_{2}+ m_{3}}{\mu_{0}}\right)}
\end{equation}
In the above equation, the value of $\alpha_s$ at $\mu_0$ = 1 GeV is considered 0.6 as shown in Table \ref{tab:table1}.
The six-dimensional hyperradial Schr{ö}dinger equation described by equation (\ref{eq:9}) has been solved in
the variational scheme with the hyper-Coloumb trial radial wave function which is given by \cite{Thakkar2011,Santopinto1998}
\begin{eqnarray}\nonumber\label{eq:12}
\psi_{\nu \gamma}=&&\left[\frac{(\nu -\gamma)! (2g)^6}{(2\nu+5)(\nu +\gamma+4)!}\right]^{\frac{1}{2}} (2 g x)^{\gamma}\\
 &&\,\,\,\,\,\,\,\,\,\,\,\,\,\,\,\,\,\,\,\,\,\,\,\,\,\,\,\,\,\,\,\,\,\,\,\,\,\,\,\,\,\,\,\,\,\,\,\,\,\,\,\,\,\,\,  \times\,\, e^{-g x} L_{\nu -\gamma} ^{2\gamma+4} (2 g x)
\end{eqnarray}
The wave function parameter g and hence the energy eigenvalue are obtained by applying virial theorem.
The baryon masses are determined by the sum of the model quark masses, kinetic
energy and potential energy as
\begin{equation}\label{eq:13}
M_B = \sum_i{m_i} + \langle H \rangle
\end{equation}

\section{Semileptonic Transition of $\Lambda_b$ $\rightarrow$ $\Lambda_c$ $\ell$ $\bar{\nu}$}\label{sec:3}
In the approximation of infinite heavy quark masses ($m_{b,c}\rightarrow\infty$), the masses of heavy quarks b and c are
much larger than the strong interaction scale $\Lambda_{QCD}$. The spin of the heavy quark decouples from light quark and
gluon degrees of freedoms. This flavour and spin symmetry provide several model independent relations for the heavy to heavy baryonic form factors. In the heavy quark limit, the six form factors $F_i$, $G_i$  $(i = 1,2,3)$ defining semileptonic transition of $\Lambda_b$ $\rightarrow$ $\Lambda_c$ $\ell$ $\bar{\nu}$ are related to a unique universal Isgur-Wise function ($\xi(\omega)$) only and they are written as

\begin{equation}
F_1(q^2)=G_1(q^2)=\xi(\omega),\,\, F_2=F_3=G_2=G_3=0
\end{equation}
where $\omega$ is the scalar invariant ($\omega$ $\equiv$ $\upsilon_{\Lambda_b}$$\,\cdot\,$$\upsilon_{\Lambda_c}$) which is related to the squared four-momentum transfer between the heavy baryons, $q^2$, by an equation
\begin{equation}
\omega=\frac{m^2_{\Lambda_b}+m^2_{\Lambda_c}-q^2}{2m_{\Lambda_b}m_{\Lambda_c}}
\end{equation}

 In the literature, various approaches exist to calculate Isgur-Wise function in absence of any standard formulation. Here, the Isgur-Wise function can be calculated using Taylor's series expansion at the zero recoil point (\,$\xi(\omega)|_{\omega=1}=1$) as
\begin{equation}\label{eq:15}
\xi(\omega)=1-\rho^2 (\omega-1)+c(\omega-1)^2+....
\end{equation}
where $\rho^2$ is the magnitude of the slope and c is the curvature (convexity parameter) of Isgur-Wise function ($\xi(\omega)$) at $\omega=1$. $\rho^2$ and c can be written as
\begin{equation}
\rho^2= - \frac{d\xi(\omega)}{d\omega}|_{\omega=1}\,\,\,\,\,\,\,\,;\,\,\,\,\,\,\,\,\,  c=\frac{d^2\xi(\omega)}{d\omega^2}|_{\omega=1}
\end{equation}

The Isgur-Wise function for the weak decay of heavy baryons transition in the HCQM can be written as an overlap integral of the baryon wave functions and has the form \cite{Hassanabadi2014}
\begin{equation}\label{eq:17}
\xi(\omega)=16\, \pi^2\int_{0}^{\infty} |\psi_{\nu\gamma}(x)|^2\, cos(p x)\, x^5 \,dx
\end{equation}
Generally, the overlap integral which involves the final and the initial wavefunction is used to calculate transition matrix elements. In the above equation, only $|\psi(x)|^2$ comes into the picture instead of the overlap integral of the final and the initial state. This is because, we have investigated the Isgur-Wise function near the zero recoil point ($\omega=1$), where the four velocities of the baryons before and after transitions are identical. Now, by expanding $cos(p x)$, we get
\begin{equation}\label{eq:18}
cos(p x)= 1-\frac{p^2x^2}{2!}+\frac{p^4x^4}{4!}+.....
\end{equation}
where $p^2$ is the square of virtual momentum transfer which can be written as $p^2=2 m^2 (\omega-1)$. After substituting Eqn.(\ref{eq:18}) into Eqn.(\ref{eq:17}) and then comparing Eqn.(\ref{eq:17}) with Eqn.(\ref{eq:15}), the slope and curvature of the Isgur-Wise function in HCQM can be derived as
\begin{equation}
\rho^2=16\, \pi^2 m^2\int_{0}^{\infty} |\psi_{\nu\gamma}(x)|^2\,  x^7 \,dx
\end{equation}

\begin{equation}
c=\frac{8}{3}\, \pi^2 m^4\int_{0}^{\infty} |\psi_{\nu\gamma}(x)|^2\,  x^9 \,dx
\end{equation}

The Isgur-Wise function mentioned by Eqn.(\ref{eq:17}) depends on the product of two terms, the first is the square of the modulus of the wave function ($|\psi_{\nu\gamma}(x)|^2$)  and the second is $cos(p x)$. The value of the Cosine term appeared in the Isgur-Wise function becomes 1, when we put $\omega=1$. And the remaining term $|\psi_{\nu\gamma}(x)|^2$ gives $\xi(\omega)=1$, while integrating it for the ground state $(\nu=0, \gamma=0)$ wave function at the zero recoil point.

Once the Isgur-Wise function is obtained, one can predict the semileptonic transition of heavy baryons. The differential decay width for the semileptonic transition of heavy baryon can be written as \cite{Guo1996}

\begin{eqnarray}\nonumber\label{eq:21}
\frac{d\Gamma}{d\omega}=\frac{2}{3}\, m^4_{\Lambda_c}\,m_{\Lambda_b}\, A && \xi^2(\omega)\,\sqrt{\omega^2-1}\\
&&\times \left[3 \omega (\eta+\eta^{-1})-2-4\omega^2\right]
\end{eqnarray}

Here, $\eta=m_{\Lambda_b}/m_{\Lambda_c}$ and $A=\frac{G^2_{F}}{(2\pi)^3}$ $|V_{cb}|^2$ $Br(\Lambda_c \rightarrow ab)$. $G_{F}$ is the Fermi coupling constant and $|V_{cb}|$ is the Kobayashi-Maskawa matrix element. $Br(\Lambda_c \rightarrow ab)$ is the branching ratio through which $\Lambda_c$ is observed.

%($\frac{d\gamma}{Ad\omega}$)

To calculate the total decay width, we have integrated the above Eqn.(\ref{eq:21}) over the solid angle as
\begin{equation}
\Gamma=\int_{1}^{\omega_{max}}\frac{d\Gamma}{d\omega} \,\,d\omega
\end{equation}
where the upper bound of the integration $\omega_{max}$ is the maximal recoil ($q^2=0$) and it can be written as
\begin{equation}
\omega_{max}=\frac{m^2_{\Lambda_b}+m^2_{\Lambda_c}}{2\,m_{\Lambda_b}m_{\Lambda_c}}
\end{equation}
where $m_{\Lambda_b}$ and $m_{\Lambda_c}$ are the masses of ${\Lambda_b}$ and ${\Lambda_c}$ baryons respectively.\\
\begin{table}[h]
\begin{center}
\caption{\label{tab:table1}Quark mass parameters (in GeV) and constants used in the calculations.}
\begin{tabular}{cccccccc}
\hline\hline
${m_{u}}$ &${m_{d}}$&${m_{c}}$&  ${m_{b}}$ & $n_{f}$ & $\alpha_s(\mu_{0}$=1 GeV)\\
%(GeV)&(GeV)&(GeV)&(GeV)&-&-&-&\\
\hline
0.330 & 0.350&1.55  &4.95  &4 &0.6\\

\hline\hline
\end{tabular}
\end{center}
%\end{ruledtabular}
\end{table}
%$A=\frac{G^2_F}{(2\pi)^3}$ $|V_{cb}|^2$ $Br(\Lambda^+_c \rightarrow ab)$
\section{Result and Discussions}
\label{sec:4}
We have chosen the quark mass parameters as $m_u$ = 0.33 GeV, $m_d$ = 0.35 GeV, $m_c$ = 1.55 GeV  and $m_b$ = 4.95 GeV (See Table \ref{tab:table1}) to calculate the masses of $\Lambda_c$ and $\Lambda_b$ baryons in the Hypercentral Constituent Quark Model (HCQM). The computed masses of $\Lambda_c$ and $\Lambda_b$ baryons are mentioned in Table \ref{tab:02}.  The calculated mass of $\Lambda_c$ baryon is 2.232 GeV and the mass of $\Lambda_b$ baryon is 5.619 which is in good agreement with the experimental results and the other model predictions.
\begin{figure}
\centering
\includegraphics[scale=0.33]{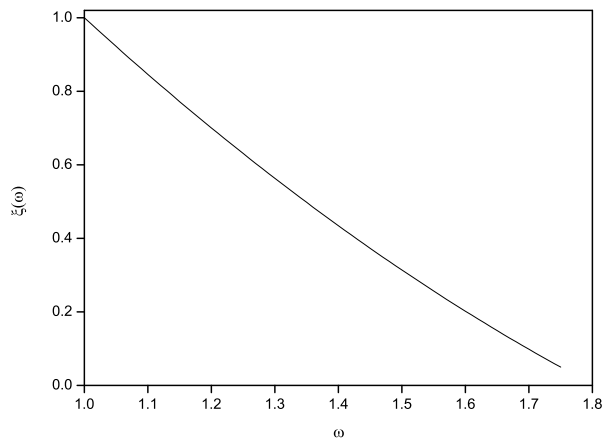}
\caption{\label{fig:1} The Isgur-Wise function ($\xi(\omega)$) for the $\Lambda_b$ $\rightarrow$ $\Lambda_c$ $\ell$ $\bar{\nu}$ semileptonic decay.}
\end{figure}

\begin{table}
\centering
%\begin{ruledtabular}
\caption{\label{tab:02}Masses of $\Lambda_{c}$ and $\Lambda_{b}$ Baryons in GeV.}
% \resizebox{\textwidth}{!}{

\begin{tabular}{ccccccccc}
\hline\hline
$M_{\Lambda_{c}}$ & Reference & $M_{\Lambda_{b}}$ & Reference\\
\hline
2.232 & This work & 5.619 & This work \\
2.286&PDG \cite{PDG2018}&5.619&PDG \cite{PDG2018}\\
2.286&\cite{Ebert2011}&5.620&\cite{Ebert2011}\\
2.285&\cite{Yoshida2015}&5.618&\cite{Yoshida2015}\\
2.286&\cite{Chen2015}&5.619&\cite{Chen2015}\\
2.268&\cite{Roberts2008}&5.619&\cite{Wei}\\
2.272&\cite{Miguraa2006}&5.612&\cite{Yamaguchi2015}\\
\hline\hline
\end{tabular}
%\end{ruledtabular}
\end{table}

\begin{table}
%\begin{ruledtabular}
\caption{\label{tab:03}Predictions for the slope at zero recoil of the baryonic Isgur-Wise function $\xi(\omega)$.}
% \resizebox{\textwidth}{!}{
\centering
\begin{tabular}{ccccccccc}
\hline\hline
Slope $(\rho^2)$ & & Approach & Reference\\
\hline
1.58&&This Work\\
1.63$\pm$0.07$\pm$0.08&&LHCb Collaboration&\cite{LHCb2017}\\
1.51&&Relativistic Quark Model&\cite{Ebert2006}\\
1.35$\pm$0.13&&QCD sum rule&\cite{Huang2005}\\
$1.2^{+0.8}_{-1.1}$&&Lattice QCD&\cite{Bowler1998}\\
1.61&& Hyperspherical &\cite{Hassanabadi2014}\\
1.3&&Large-$N_c$ Limit&\cite{Jenkins1993}\\
2.4&&MIT Bag Model&\cite{Sadzikowski1993}\\
1.4-1.6&&Bethe-Salpeter Equation&\cite{Guo1996}\\
1.47&& Light-front approach  &\cite{Hong-Wei2008}\\
1.5&&spectator quark model&\cite{Hussain1991}\\
2.03$\pm$$0.46^{+0.72}_{-1.00}$&& DELPHI collaboration&\cite{DELPHI Collaboration2004}\\

\hline\hline
\end{tabular}
%\end{ruledtabular}
\end{table}

The behaviour of the variation of Isgur-Wise function with respect to $\omega$ is shown in Fig. \ref{fig:1}. The slope $(\rho^2)$ at zero recoil of the baryonic Isgur-Wise function $\xi(\omega)$ is computed and the result along with the other theoretical predictions are listed in Table \ref{tab:03}. The calculated value for the slope at zero recoil of the baryonic Isgur-Wise function is 1.58 which fairly agrees with other theoretical predictions within the theoretical errors. The result obtained from the relativistic quark model \cite{Ebert2006} for slope  of the Isgur-Wise function is $1.51$ which indicates good agreement with our prediction. The predicted value for the slope of the Isgur-Wise function is in accordance with the experimental value 1.63$\pm$0.07$\pm$0.08, recently reported by LHCb collaboration \cite{LHCb2017}. The overall range of the slope predicted by all the theoretical predictions varies from 1.2 to 1.61.
 The Spectator quark model \cite{Hussain1991} has predicted the relation between the slope of baryonic Isgur-Wise function ($\rho^2_{B}$) and the slope of mesonic Isgur-Wise function ($\rho^2_{M}$) through $\rho^2_{B}$ = 2 $\rho^2_{M}$ - 1/2. The Heavy Flavor Averaging Group \cite{Heavy Flavor Averaging Group2014} has reported $\rho^2_{M}$ $\approx$ 1. After submitting this slope of mesonic Isgur-Wise function, the value obtained for $\rho^2_{B}$ is 1.5. Thus the slope of the Isgur-Wise function obtained from the Spectator quark model \cite{Hussain1991} is in agreement with the calculated value for the slope of the Isgur-Wise function in this paper. By comparing the slope of the Isgur-Wise function at the zero recoil point for the  heavy baryon and the heavy meson, we predict that the Isgur-Wise function for the baryons should be a much steeper function of $\omega$ than the corresponding function for mesons. Our computed value of convexity parameter c is 0.42. The other theoretical model (reference \cite{Hassanabadi2014}) has predicted the value of convexity parameter c = 0.56 which is comparatively higher than our prediction.
\begin{figure}
\centering
\includegraphics[scale=0.33]{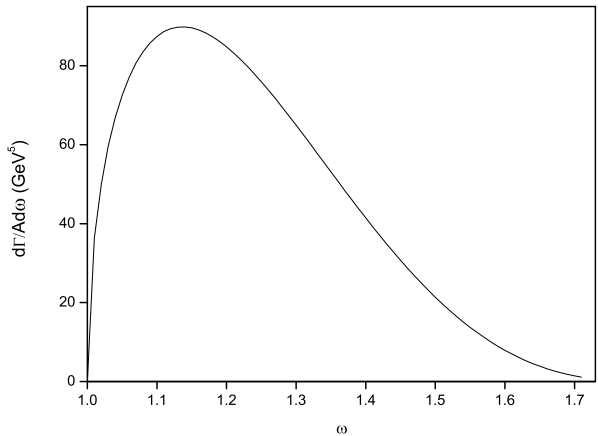}
\caption{\label{fig:2}The variation of differential decay rate  for the $\Lambda_b$ $\rightarrow$ $\Lambda_c$ $\ell$ $\bar{\nu}$ semileptonic decay. }
\end{figure}
\begin{table}
\caption{\label{tab:04}Comparison of theoretical predictions for the $\Lambda_b$ $\rightarrow$ $\Lambda_c$ $\ell$ $\bar{\nu}$ semileptonic decay parameters with available experimental data.}
\centering
\begin{tabular}{cccrccccc}
\hline\hline
Decay Width $\Gamma$ &Reference& Branching &Reference\\
 (in $10^{10} s^{-1})$	&		&Ratio	Br$(\%)$	&		\\
\hline
4.11	&	This Work	&6.04		&	This Work	\\
3.52	&	\cite{Azizi2018}	&	6.04$\pm$1.70	&	\cite{Azizi2018}	\\
5.02	&	\cite{Ebert2006}	&	6.9	&	\cite{Ebert2006}	\\
4.42	&	\cite{Faustov2016}	&	6.48	&	\cite{Faustov2016}	\\

4.86	&\cite{Gutsche2015}&	6.9	&	\cite{Gutsche2015}	\\
2.15$\pm$0.08$\pm$0.11	&	\cite{Detmold2015}	&	4.83	&	\cite{Dutta2016}	\\
5.39	&	\cite{Ivanov1997}	&	6.2$^{+1.4}_{-1.3}$	&Expt.\cite{PDG2018}	\\
3.52$^{+2.2}_{-1.9}$	&	\cite{Bowler1998}	& $5.0^{+1.1+1.6}_{-0.8-1.2}$		&	 Expt.\cite{Abdallah2004}	\\
5.9	&	\cite{Singleton1991}	&		&		\\
4.92	&	\cite{Hassanabadi2014}	&		&		\\
4.2 -- 5.7&\cite{Guo1996}\\
5.14&\cite{korner1994}\\
5.1&\cite{Cheng1996}\\
6.09&\cite{Ivanov1999}\\
5.08$\pm$1.3&\cite{Cardarelli1999}\\
5.82&\cite{Albertus2005}\\
5.39	&	\cite{Pervin2005}\\
\hline\hline
\end{tabular}
%\end{ruledtabular}
\end{table}

We are able to calculate the decay width and the branching ratio of $\Lambda_b$ $\rightarrow$ $\Lambda_c$ $\ell$ $\bar{\nu}$ semileptonic decay from the obtained Isgur-Wise function. The plot for differential decay width is shown in Fig. \ref{fig:2}. The experimental value of $m_{\Lambda_b} = 5.619$ GeV and $m_{\Lambda_c} = 2.286$ GeV (PDG \cite{PDG2018}) are used to calculate $\Lambda_b$ $\rightarrow$ $\Lambda_c$ $\ell$ $\bar{\nu}$ semileptonic decay. Table \ref{tab:04} provides a comparison of theoretical predictions for the $\Lambda_b$ $\rightarrow$ $\Lambda_c$ $\ell$ $\bar{\nu}$ semileptonic decay parameters with available experimental data. While comparing our result for decay width with other theoretical predictions, we have converted the GeV unit to $s^{-1}$ unit in some predictions.  Our calculated result for the semileptonic decay width of $\Lambda_b$ baryon is $4.11 \times 10^{10} s^{-1}$. From Table \ref{tab:04}, we see that the decay widths from different theoretical predictions vary from 2.15$\times 10^{10} s^{-1}$ to 6.09$\times 10^{10} s^{-1}$. The relativistic quark model \cite{Faustov2016} has predicted the value of semileptonic decay width $\Gamma$ = $4.42 \times 10^{10} s^{-1}$ which is in good agreement with our computed result.  Our calculated branching ratio for the $\Lambda_b$ $\rightarrow$ $\Lambda_c$ $\ell$ $\bar{\nu}$ semileptonic decay is 6.04 $\%$. We have used mean life time $\tau_{\Lambda_b} = 1.47 \times 10^{-12} s$ and the value of $|V_{cb}| = 0.041$ as given in PDG \cite{PDG2018} to calculate the branching ratio. The computed result for the branching ratio nicely agrees with the average experimental value (6.2$^{+1.4}_{-1.3}$ $\%$) within its experimental error reported by PDG \cite{PDG2018}.\\
\section{Conclusions}
\label{sec:5}
The transition properties for $\Lambda_b$ $\rightarrow$ $\Lambda_c$ $\ell$ $\bar{\nu}$ semileptonic decay are studied within the framework of a Hypercentral Constituent Quark Model. After fixing the model parameters using the ground state mass of $\Lambda_b$ baryon, the slope at zero recoil of the baryonic Isgur-Wise function is computed. With the help of the Isgur-Wise function, the exclusive semileptonic decay width and the branching ratio of $\Lambda_b$ baryon are calculated. The computed results for $\Lambda_b$ $\rightarrow$ $\Lambda_c$ $\ell$ $\bar{\nu}$ semileptonic decay and its branching ratio are in agreement with available experimental observations and with other model predictions. The HCQM gives plausible predictions for the Isgur-Wise function, decay width and the branching ratio corresponding to the $\Lambda_b$ $\rightarrow$ $\Lambda_c$ $\ell$ $\bar{\nu}$ semileptonic decay.\\\\
\textbf{Acknowlwdgements}\\\\
The author would like to thank the referee for his valuable comments and suggestions to improve the present paper.
%\textbf{References}

\end{document}